\documentclass[preprint]{aastex631}
\usepackage{amsmath}
\usepackage{graphicx}
\graphicspath{{./figs/}}
\usepackage{physics}
\usepackage{lipsum}
\usepackage[version=4]{mhchem}
\usepackage{pifont}
\let\tablenum\relax
\usepackage[retain-unity-mantissa=false]{siunitx}[=v2]

\usepackage{printlen}
\usepackage{layouts}
\usepackage{multirow}

\submitjournal{ApJ}
\begin{document}
\title{The Runaway Greenhouse Effect on Hycean Worlds}

\author[0000-0001-5271-0635]{Hamish Innes}
\affiliation{Atmospheric, Oceanic and Planetary Physics, Department of Physics, University of Oxford, UK}
\author[0000-0002-8163-4608]{Shang-Min Tsai}
\affiliation{Atmospheric, Oceanic and Planetary Physics, Department of Physics, University of Oxford, UK}
\affiliation{Department of Earth Sciences, University of California, Riverside, California, USA}
\author[0000-0002-5887-1197]{Raymond T. Pierrehumbert}
\affiliation{Atmospheric, Oceanic and Planetary Physics, Department of Physics, University of Oxford, UK}

\correspondingauthor{Hamish Innes}
\email{hamish.innes@physics.ox.ac.uk}
\begin{abstract}
Hycean worlds are a proposed subset of sub-Neptune exoplanets with substantial water inventories, liquid surface oceans and extended hydrogen-dominated atmospheres that could be favourable for habitability. In this work, we aim to quantitatively define the inner edge of the Hycean habitable zone using a 1D radiative-convective model. As a limiting case, we model a dry hydrogen-helium envelope above a surface ocean. We find that 10 to 20 bars of atmosphere produces enough greenhouse effect to drive a liquid surface ocean supercritical when forced with current Earth-like instellation. Introducing water vapour into the atmosphere, we show the runaway greenhouse instellation limit is greatly reduced due to the presence of superadiabatic layers where convection is inhibited. This moves the inner edge of the habitable zone from $\approx 1$ AU for a G-star to 1.6 AU (3.85 AU) for a Hycean world with a \ce{H2}-\ce{He} inventory of 1 bar (10 bar). For an M-star, the inner edge is equivalently moved from 0.17 AU to 0.28 AU (0.54 AU). Our results suggest that most of the current Hycean world observational targets are not likely to sustain a liquid water ocean. We present an analytical framework for interpreting our results, finding that the maximum possible OLR scales approximately inversely with the dry mass inventory of the atmosphere. We discuss the possible limitations of our 1D modelling and recommend the use of 3D convection-resolving models to explore the robustness of superadiabatic layers.
\end{abstract}
\section{Introduction}

\subsection{Sub-Neptunes and Hycean Worlds}
Theorists have long predicted the existence of small, water-rich planets forming outside the ice line with no solar system analogue \citep{kuchner2003,leger2004}. With the launch of the Kepler space telescope, the population of detected sub-Neptunes increased by orders of magnitude \citep{Batalha2014}. They are considered one of the most abundant types of planet in the galaxy (alongside super-Earths) \citep{Petigura2013, Marcy2014b,Winn2015}. Limited data leaves their inner compositions largely unconstrained, with structures ranging from almost pure water to rocky/iron cores producing the same mass and radius when given a hydrogen-dominated envelope \citep{Rogers2010, rogers2011, Marcy2014b, Lopez2012, Rogers2015, Madhusudhan2020}. Population demographics provide insights into the composition of smaller planets. The discovery of the radius valley \citep{Fulton2017} separating the population of smaller super-Earths and larger sub-Neptunes provides a way to distinguish these two types of planet. By studying how the position of the valley varies with instellation, there is consensus that it is shaped by atmospheric loss, either via core-powered mass loss \citep{gupta2019} or photoevaporative loss \citep{Owen2013}. 

``Hycean Worlds" \citep{madhusudhan2021} are a proposed subset of water-rich sub-Neptunes with hydrogen-dominated atmospheres. The sub-Neptune K2-18 b \citep{Cloutier2017,Cloutier2019} has been the subject of multiple studies owing to the claimed detection of water vapour in its atmosphere \citep{Benneke2019c,Tsiaras2019}, with \cite{Madhusudhan2020} showing that its mass and radius were consistent with being a Hycean world. Further work on the habitability \citep{madhusudhan2021} and interior structure \citep{nixon2021} of Hycean worlds confirm there is a wide range of mass-radius parameter space potentially occupied by this type of planet. Their larger radii and masses compared to terrestrial-type habitable planets makes them more amenable to observational characterisation, with several temperate candidates having already been identified \citep[see Table~1 of][]{madhusudhan2021}.

\subsection{Planetary Habitability and the Runaway Greenhouse Effect}
\subsubsection{Background}
Most models of planetary habitability focus on atmospheres where the mean molecular weight (MMW) of the background atmosphere is greater than the condensible component (and for good reason -- such is the case on Earth). Pioneering work related to Venus's atmosphere \citep{Simpson1929, Komabayasi1967,ingersoll1969, nakajima1992} established a maximum instellation above which a planet can no longer increase its infrared cooling into space. This is known as the ``runaway greenhouse" limit \citep{ingersoll1969}. As high surface temperatures make water vapour a non-negligible component of the atmosphere, there comes a point where the atmosphere becomes optically thick to infrared radiation, decoupling the surface temperature from a fixed photospheric temperature. If a planet's instellation exceeds this value, its water inventory evaporates into the atmosphere, with photodissociation and atmospheric escape eventually driving it into space. The inner edge of the habitable zone (HZ) is typically calculated assuming the runaway greenhouse effect is the limiting factor influencing habitability \citep{Kasting1993, Kopparapu2013}. 

\subsubsection{The Impact of an \ce{H2}-\ce{He} Background and Convective Inhibition} 
As interest around hydrogen-dominated habitable worlds has grown, it is important to assess the impact of hydrogen as a background gas as opposed to nitrogen or other high MMW gas. Hydrogen gas's main source of opacity comes from its collision-induced absorption (CIA) spectrum. Since the strength of absorption scales as the density of the atmosphere squared, CIA becomes an important factor in \ce{H2} atmospheres of around 1 bar or thicker, with 40 bars of pure \ce{H2} allowing habitable surface temperatures out to 1.5 AU for an M star and 10 AU for a G star \citep{Pierrehumbert2011}. Accounting for absorption and water vapour feedbacks is necessary to further constrain the inner edge of the habitable zone. In \cite{madhusudhan2021}, inner edge calculations are, for the most part, made assuming a constant water mixing ratio of 10\% (or less, if condensation occurs). Calculations with a water-saturated atmosphere have also been performed \citep[e.g., Figure 9 of][]{piette2020a}, however their model neglects compositional effects (see below). When hydrogen is used as the background gas in classical runaway greenhouse calculations \citep{Koll2019}, the presence of \ce{H2} introduces novel behaviour on account of its low MMW. Unlike for high MMW background gases which tend to raise the runaway greenhouse limit and lead to non-monotonic changes in outgoing longwave radiation (OLR) with surface temperature, \ce{H2} atmospheres monotonically approach the pure steam limit with increasing surface temperature.

Our work aims to build on \cite{Koll2019} by introducing the effect of MMW-induced convective inhibition on the temperature profiles of hydrogen-dominated planets. For atmospheres with a condensing component heavier than the background gas, decreasing temperatures with altitude leads to a sharp decrease in MMW between the lower atmosphere and the upper atmosphere. If the concentration of the condensible is high enough, compositional gradients stabilise the atmosphere to convection \citep{Guillot1995,Li2015,Leconte2017} and double-diffusive instabilities \citep{Leconte2017} even if the lapse rate is super-adiabatic. Molecular gradient-induced convective inhibition has been invoked to explain periodic storms in Saturn's atmosphere \citep{Li2015}, stable wave ducts for gravity waves in Jupiter's atmosphere \citep{Ingersoll1995} and explain the step-like behaviour of methane's abundance in Uranus's condensation layer \citep{irwin2022}. Neglecting this effect results in an underestimation of the deep atmospheric temperature in gas giant planets \citep{Leconte2017}. Recent studies have investigated convective inhibition due to condensing silicates in a \ce{H2}-\ce{He} envelope, which alters the radius and cooling times of young super-Earths and sub-Neptunes \citep{Misener2022, Markham2022}
    
The atmospheric structure of runaway atmospheres is often assumed to be on a moist adiabat integrated upwards from the surface. However, if the surface temperature is high enough, high moisture contents will inhibit convection and lead to radiative layers in the lower atmosphere. We will discuss this effect in further detail in Section~\ref{sec:h2h2o}. Our aim is to calculate new limiting instellations for sub-Neptunes hosting a liquid water ocean with a hydrogen-dominated atmosphere.

\subsubsection{Super-Runaway States}
Even a liquid water ocean can be too hot to be habitable for life as we know it, but from the standpoint of planetary structure a second important transition occurs when the surface temperature of the water layer reaches the critical point. At this point, the liquid-gas phase transition of water disappears. More importantly for our purposes, \ce{H2} is completely miscible with supercritical water \citep{soubiran2015}, so that gaseous solubility is no longer limited by Henry's Law (miscibility is expected for helium and other gases as well.)  In consequence, a hydrogen envelope could not remain distinct from the supercritical water interior, but instead would mix into it and be diluted into the supercritical ocean. It is an interesting theoretical question how long it would take for such mixing to occur if the planet started out with a distinct hydrogen envelope, but the more plausible scenario for a planet with an initial \ce{H2}-\ce{H2O} composition is that the system would start out with a supercritical \ce{H2}-\ce{H2O} mixture. The \ce{H2} would never phase-separate into a distinct layer if the equilibrium radiation balance maintains supercritical conditions with respect to water.  For this reason, we put particular emphasis on defining planetary parameters for which the ocean surface temperature approaches the critical point of water. The chief goal of this paper is to identify conditions in which a subcritical liquid water ocean can coexist with a hydrogen-rich atmosphere, though we do offer some speculations on the state a sub-Neptune settles into when those conditions are not met. 

\subsection{Paper Structure}
The paper is structured as followed. In Section~\ref{sec:h2he} we will discuss the \ce{H2}-\ce{He} inventory required to produce a strong enough greenhouse effect to drive surface temperatures to the critical point of water. This will give an upper-bound on the instellation a planet can receive from its host star before entering a runaway state. In Section~\ref{sec:h2h2o} we extend our model to deal with atmospheres with a water vapour component. We will then discuss our results in the context of the literature in Section~\ref{sec:discussion}.

\section{Hydrogen-helium atmosphere above a water ocean}\label{sec:h2he}
\subsection{Model}
To find an upper bound on the hydrogen inventory of a planet that maintains a liquid water surface, we first model a hydrogen-helium atmosphere above a water ocean. We assume the temperature structure of the atmosphere is the dry adiabat for a hydrogen-helium mixture with solar abundances \citep{Asplund2009}, neglecting the contribution of water vapour to the atmosphere (similar to the calculation in \cite{Pierrehumbert2011}). This is an unphysical assumption given warm temperatures will naturally lead to the water evaporating from the surface ocean and mixing with the hydrogen-helium gas. However, it serves as an easily calculated upper limit on the hydrogen-helium content of a planet with temperate instellation since we expect the addition of water vapour to act as a greenhouse gas, reducing the ability of the planet to cool. Our choice of a purely adiabatic atmosphere is justified from test runs using a full radiative-convective iteration (see Section~\ref{sec:h2h2o}). We found that the tropopause (at $\approx$ 0.05 bars) was typically at lower pressures than the infra-red photosphere (at $>0.1$ bar).

Radiative fluxes were calculated in the longwave (LW) and shortwave (SW) regions of the spectrum using the SOCRATES radiative transfer code \citep{Edwards1996}, the details of which can be found in Appendix~\ref{sec:app_radflux}. In the SW calculation, a surface albedo of 0.12 is specified \citep{goldblatt2013}.

Since we are interested in calculating the maximum hydrogen inventory possible before an ocean is driven supercritical, we calculate fluxes for an atmosphere on a dry adiabat with surface temperature equal to the critical temperature of water, \SI{647}{K}. We vary the surface pressure logarithmically between 0.5 bar and the critical pressure of water, 220 bar. The SW fluxes were calculated using a solar instellation value of \SI{1361}{\W\per\m\squared}, however we note that since the temperature structure of the atmosphere is fixed, we can simply multiply our results by a constant factor to retrieve the SW fluxes for any arbitrary instellation. The surface gravity was kept constant at the Earth value, \SI{9.81}{\m\per\s\squared}, throughout. We consider cases where the incoming stellar spectrum has that of a G star or an M star (details of which are found in Appendix~\ref{sec:app_radflux}). 

\subsection{Results}
\begin{figure}[h]
    \centering
    \includegraphics[width=0.55\linewidth]{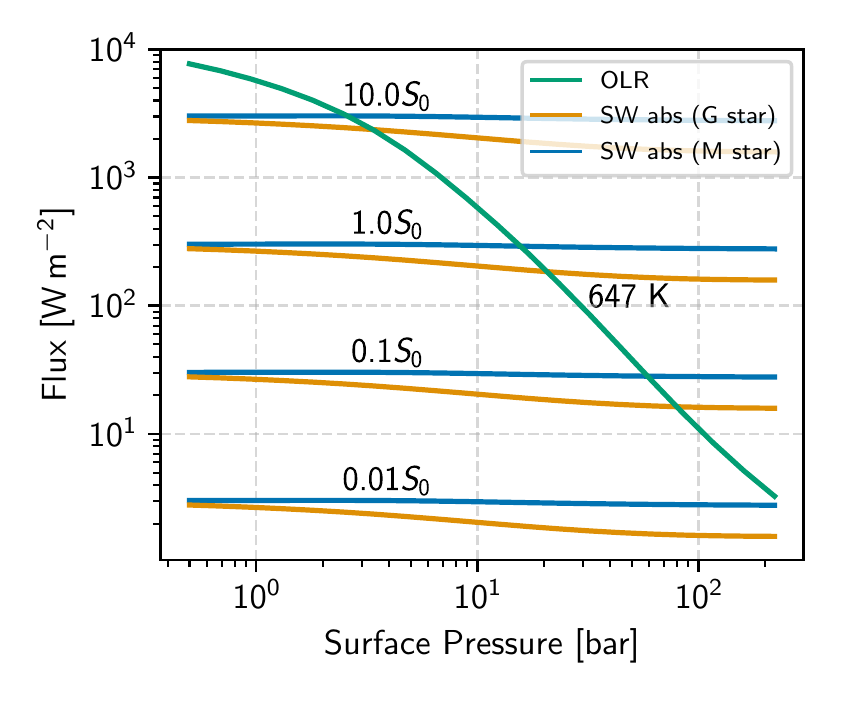}
    \caption{OLR and SW absorption as a function of surface pressure. The green curve represents the OLR of an \ce{H2}-He atmosphere on a dry adiabat with surface temperature equal to the critical temperature of water. The blue and orange curves show the SW absorption at different instellations relative to solar ($S_0$) for an M star and G star respectively.}
    \label{fig:h2he}
\end{figure}

Only moderate pressures of hydrogen and helium are required to force surface temperatures to supercritical values. Figure~\ref{fig:h2he} shows the OLR and SW absorption of these atmospheres as a function of surface pressure. The intersection points between the OLR curve (green) and the SW absorption curves (blue and orange depending on stellar type) signify atmospheric configurations in global equilibrium. For a pure \ce{H2}-He atmosphere with solar instellation, $S_0$, approximately 10 bars of atmosphere will cause a large enough greenhouse effect due to CIA to drive the surface ocean supercritical. The SW absorption in the G-star experiments was reduced relative to the M-star thanks to the enhanced Rayleigh scattering cross sections at low wavelengths. However, the change in stellar type causes minimal differences in the qualitative behaviour of our model. Naturally, lower instellation values require larger atmospheric masses to warm the surface to the critical temperature, and atmospheres irradiated with less than 1\% of solar radiation would require a surface pressure greater than the critical pressure of water to reach the critical temperature.

We note the surface gravity in these experiments was held constant at Earth's value. As noted in \cite{Pierrehumbert2011}, the OLR depends on the surface gravity in the combination of $p_s^2/g$, so changing $g$ can be accounted for simply by rescaling $p_s$. From Figure~\ref{fig:h2he} we see that increasing the surface gravity from its terrestrial value would be equivalent to lowering the surface pressure, i.e. increasing the OLR (assuming the change in Rayleigh scattering, which scales as $p_s/g$, does not cause the planetary albedo to change significantly). Therefore, for the same level of instellation, a planet with a higher surface gravity can host a higher pressure atmosphere before going supercritical. Nevertheless, we expect the surface gravity of temperate sub-Neptunes to be of a similar order of magnitude to the Earth. The four temperate sub-Neptunes with constrained mass and radius \citep[Table 2 of ][]{pierrehumbert2023} have surface gravities spanning the range \SIrange{4}{24}{\m\per\s\squared}.

\section{Adding water vapour to the atmosphere}\label{sec:h2h2o}
\subsection{Model Setup}
\subsubsection{Thermodynamics}
Water is added to each layer at its saturation pressure value (i.e. the relative humidity is assumed to be 100\%). We use the analytic expressions for the saturation vapour pressure given by \cite{Wagner2002}, generating a lookup table of values for each temperature for faster on-the-fly calculation. 
\subsubsection{Radiative Procedures}
As in Section~\ref{sec:h2he}, radiative fluxes are calculated with the SOCRATES radiative transfer code. Unlike in Section~\ref{sec:h2he} where an inverse modelling approach was taken, we now timestep the model by iterating the temperature according to:
\begin{equation}
    \pdv{T}{t} = A(p) \pdv{F_{\text{net}}}{p}
\end{equation}
where $T$ is the temperature, $F_{\text{net}}$ is the net flux two-stream flux calculated at the edge of each model layer (defined such that $F_{\text{net}}$ is positive if the net flux is upwards, in the $-p$ direction). The coefficient $A(p)$ varies in each model layer in order to approach radiative equilibrium (where 
 $\partial_p F_\text{net} = 0$) as fast as possible. We use the method described in \cite{Malik2017}, where $A(p)$ is decreased if temperature oscillations are detected in that temperature layer and increased on each iteration otherwise. For our purposes we only need to compute the correct equilibrium profile, and do not need to compute the actual time series as the system adjusts to equilibrium. 

 Using this approach, the independent variable is the instellation, $S$, as opposed to the surface temperature, $T_s$, which is permitted to vary until the atmosphere reaches local and global radiative equilibrium.
 
\subsubsection{Convective Adjustment}\label{subsubsec:conv_adj}

 After each iteration, the temperature-pressure profile is checked for instability to convection. We use the instability criterion modified for use in atmospheres where water vapour is at saturation with a lower mean molecular weight background gas \citep{Leconte2017}:
\begin{subequations}
\begin{align}
 (\nabla_{\text{am}} - \nabla_{\text{ad}})(1 - \beta q \varpi )    > 0 & \label{eq:crit}\\
\text{with}\quad \nabla_x \equiv \dv{\ln T_x}{\ln p}, &\\
 \beta \equiv \dv{\ln p_{\text{sat}}}{\ln T}, &\\
 \varpi \equiv (1 - \mu_d/\mu_v). &
\end{align}
\end{subequations}
The ``am" and ``ad" suffixes denote ``ambient" and ``adiabatic" respectively, $p_{\text{sat}}$ is the saturation vapour pressure and $\mu_d$ and $\mu_v$ are the MMWs of the dry and condensing vapour gases respectively. The first term in Equation~\ref{eq:crit} is the Schwarzschild criterion \citep{schwarzschild1906} for convective instability when the ambient lapse rate is greater than the moist adiabatic lapse rate. The second term represents the effect of mean molecular weight gradients in an environment where water vapour is held at its saturation vapour pressure value. When the water vapour concentration, $q$, exceeds a critical value, $q_c \equiv 1/\beta\varpi$, the parcel is no longer unstable to convection despite any super-adiabatic lapse rates. The value of the moist adiabatic lapse rate, $\nabla_{\text{ad}}$ is calculated using the expression in \cite{ding2016}, which accounts for water vapour being a non-negligible component of the atmosphere but neglects the effects of retained condensates. We replace occurences of $L/R_vT$ in the lapse rate formula with the more appropriate $\beta$ factor obtained directly from a lookup table, and any other factors of $L$ are calculated using a lookup table using fits from \cite{Wagner2002}. This accounts for vanishing $L$ as $T$ approaches the critical temperature of water (\SI{647}{K}). However, we note that at temperatures approaching the critical point of water, the non-idealness of water vapour will become important but is neglected in this current work for simplicity. Convective adjustment is performed pairwise on layers working upwards from the bottom of the atmosphere. If Equation~\ref{eq:crit} is satisfied, we adjust the temperature of both layers to the moist adiabat. This procedure is repeated until convergence is reached. 

Note that when $q>q_c$, a saturated layer with lapse rate $\nabla_{\text{am}}$ {\it less than} the adiabatic lapse rate -- even an isothermal layer -- is unstable, and in our scheme is convectively adjusted to the saturated adiabat; radiative cooling may then cause further steepening of the lapse rate. 

The implementation of Equation~\ref{eq:crit} leads to two radiative zones forming in the atmosphere when temperatures near the surface are high enough -- a traditional stratosphere in the upper atmosphere with low lapse rates and a moisture-inhibited radiative zone below the first convective region. Since the atmosphere is optically thick in this region, the radiative lapse rate is usually high and requires higher vertical resolution. We ran the model with 200 layers, with the bottom 100 layers dedicated to resolving this radiative region of the atmosphere. In the traditional stratosphere we implement a cold trap for moisture, setting the moisture concentration to its minimum value at pressures lower than the moisture minimum. 

\subsubsection{Surface Physics}
At the surface we implement a rudimentary heat transfer scheme which transfers sensible heat from the surface to the lowest model layer. This prevents the surface temperature becoming unphysically hotter than the lowest model layer and is a $0^{\text{th}}$ order attempt to represent some of the heat exchange processes in the turbulent boundary layer. The surface energy equation is:
\begin{equation}
\label{eq:surfaceheat}
\rho_s c_s h \dv{T_s}{t} = -F_{\text{net}}(p_s)+ c_p\rho_{\text{air}}C_D U (T(p_s)  - T_s)
\end{equation}
where values for the surface density ($\rho_s$) and heat capacity ($c_s$) are taken to be those of liquid water and $h$ (the bucket depth) is taken as \SI{1}{m}. The drag coefficient $C_D$ and the characteristic drag velocity $U$ are taken to be 0.001 and \SI{10}{\m\per\s} respectively, from \cite{Pierrehumbert2010}. The effect of varying these parameters had a negligible effect on the overall temperature structure of the atmosphere and the final conclusions of this work. 

In traditional calculations of the runaway greenhouse limit using an inverse modelling approach \citep[e.g.][]{Kasting1993}, the surface pressure $p_s$ is taken to be the sum of a constant dry component, $p_0$, and the saturation vapour pressure at the specified surface temperature $T_s$. Surface pressure can increase with temperature, allowing an increase in total atmospheric mass as more water vapor is added to the atmosphere. However, holding $p_0$ fixed as temperature increases leads to an implied change in the dry mass of the atmosphere as temperature changes, because the mean molecular weight of the atmosphere varies and the concentration of the dry mass is not uniform over the profile.  For inverse climate modeling, it is straightforward to allow $p_s$ to vary with temperature, but with radiative-convective calculations that compute equilibria using time-stepping or related iterations, the re-gridding needed to allow $p_s$ to vary becomes unwieldy and can lead to numerical issues. For that reason, we make the additional simplification of holding $p_s$ itself fixed, until conditions make such an assumption physically inconsistent. The expedient of holding $p_s$ fixed was also used in the calculations of Figure 9 in \cite{piette2020a}, which were carried out with a water-saturated atmosphere. In such calculations, there is an additional reduction in implied dry mass as temperature increases, as $p_0$ needs to be reduced in order to compensate for the increase with temperature of surface water vapor partial pressure.  As long as this change isn't drastic, it is of little consequences for our purpose, as we are not attempting to track the actual time evolution of an atmosphere. It only means that the dry air mass in the equilibrium state is somewhat different from what was specified for the initial condition of the calculation. The magnitude of the adjustment will be quantified in Section \ref{subsec:drymass}. 

When the surface saturation vapor pressure approaches the specified $p_s$, $p_0\rightarrow 0$ and it is no longer possible to keep surface pressure fixed as surface temperature is further increased. To deal with this case, we introduce a pure-steam layer for $p>p_s$, extending to a greater surface pressure $p_s'$; we then time-step the radiative-convective model only for $p<p_s$, computing the radiation from the pure steam layer assuming it to lie on the pure-steam (i.e dewpoint) moist adiabat, as discussed in \cite{Pierrehumbert2016}. Since water vapour is heavy and non-buoyant in a hydrogen-dominated background atmosphere, it will remain at the bottom of the atmosphere.   

A radiative layer in the lower levels of the atmosphere tends to dry out the upper atmosphere (due to its steep lapse rates), leading to a layered structure with pure water vapor at the bottom, nearly pure hydrogen-helium at the top, and a sharp transition layer between the two.  All of the mass of hydrogen and helium resides in the upper layer, and the opacity of this layer (which in turn depends on the dry hydrogen-helium mass) strongly affects the conditions for the surface temperature to enter a runaway state. 

In cases which require a pure steam layer at the bottom, we set the bottom of the atmosphere to a fixed surface temperature (which fixes the surface pressure as the saturation vapour pressure at this temperature). The approach in this case is to determine the instellation which is compatible with the specified surface temperature, requiring multiple runs of the time-stepped model. Unlike the usual inverse climate modeling approach, the required instellation cannot be determined by just computing the OLR corresponding to a given $T(p)$ profile, since the profile is affected by the instellation through stellar absorption within the atmosphere.  Instead, we need to guess an instellation, time-step the model (subject to a lower boundary condition provided by the steam layer) until $T(p)$ reaches equilibrium, and then check the top of atmosphere balance. The instellation is then adjusted until top of atmosphere balance is achieved.  When the surface temperature is too high, the OLR becomes decoupled from surface temperature owing to the optically thick steam layer, and so equilibrium cannot be reached when instellation exceeds a threshold value, which defines the runaway condition. For instellation above the runaway threshold, the temperature increases until some process intervenes to allow OLR to increase again, as discussed in \cite{Boukrouche2021, pierrehumbert2023}. In this paper, we do not compute the super-runaway equilibrated state, but for a water-rich sub-Neptune with a deep water layer it is certain to be above the critical point of water. 

\subsection{Experimental Procedure}

To model the effect of different dry mass paths, we run the model with two surface pressures -- 1 bar and 10 bar. As described above, this surface pressure is held constant until $q(T_s,p_s) = 1$ is reached at the bottom of the atmosphere, after which the surface temperature and pressure are increased on the pure steam adiabat. Since steep radiative lapse rates keep the atmosphere relatively dry at all pressures except very close to the surface, this acts to keep the dry mass path of the atmosphere relatively constant at $\approx$\SI{1e4}{\kg\per\m\squared} for the 1 bar initial condition and $\approx$\SI{1e5}{\kg\per\m\squared} for the 10 bar case (we will quantify this in Section~\ref{subsec:drymass}). Extending the atmosphere on the pure steam adiabat does not add any dry mass to the atmosphere (and instead assumes that the extra mass comes from evaporation from the surface ocean). We will refer to these two cases as the ``1 bar" and ``10 bar" cases interchangeably with ``\SI{1e4}{\kg\per\m\squared}" and ``\SI{1e5}{\kg\per\m\squared}" cases.

For both the 1 bar and 10 bar atmospheres, we perform separate runs with the G-star and M-star spectra described in Appendix~\ref{sec:app_radflux}. For each set of runs, we begin with a low instellation that gives surface temperatures between \SIrange[range-phrase=~--~]{270}{300}{K} and run the model to radiative-convective equilibrium. After the equilibrium state is found, we reinitialise the model with an incrementally higher instellation, chosen as a balance between numerical stability and computational efficiency. Increments of \SI{5}{\W\per\m\squared} and \SI{1}{\W\per\m\squared} were used for the 1 bar and 10 bar runs respectively. The initial temperature profile of the first run is a dry adiabat with an isothermal stratosphere -- runs at higher instellations are then initialised on the final temperature profile of the previous run. When $q(T_s,p_s)=1$ is reached at the bottom of the atmosphere (i.e. $p_{\text{sat}}(T_s)=p_s$), we switch to the procedure described above where the surface temperature is held fixed and the instellation is iterated until both local and global radiative equilibrium are attained. We run the model until the surface temperature reaches 600 K, giving us a range of surface temperatures between \SIrange[range-phrase=~--~]{270}{300}{K} and \SI{600}{K}.

\subsection{Results}\label{subsec:resultsh2h2o}
\subsubsection{Temperature and Humidity Profiles}
\begin{figure}[h]
    \centering
    \includegraphics[width=\linewidth]{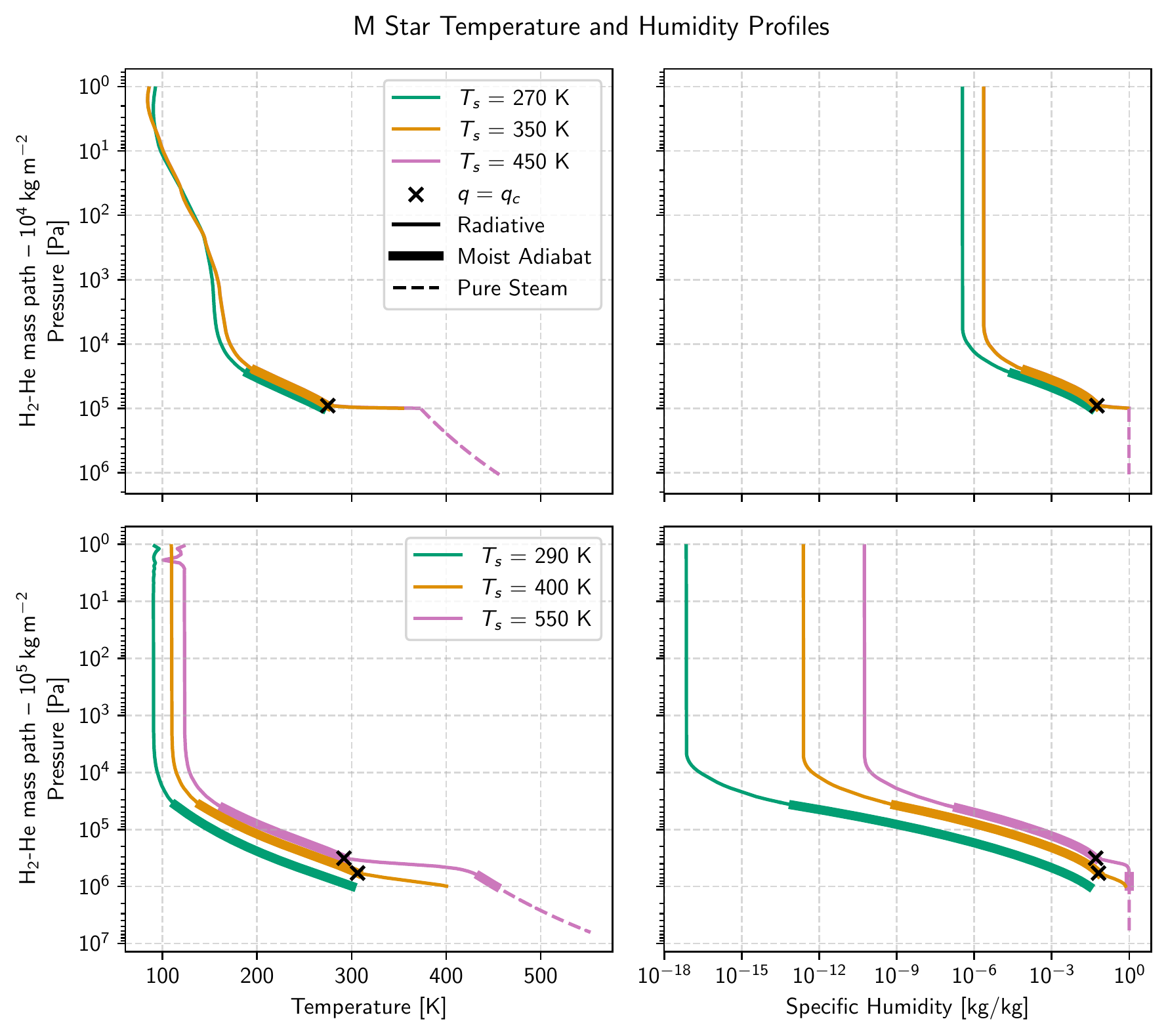}
    \caption{Sample of temperature-pressure profiles (left column) and specific humidity profiles (right column) for the M star experiments with dry mass paths of $10^4$ \si{\kg\per\m\squared} (top row) and $10^5$ \si{\kg\per\m\squared} (bottom row). The introduction of radiative layers in the lower atmosphere causes a sharp increase in the surface temperature before the lower atmosphere becomes pure steam. The radiative layers have lower lapse rates in the $10^5$ \si{\kg\per\m\squared} case because less SW radiation penetrates to the lower atmosphere.}
    \label{fig:tp_mstar}
\end{figure}

\begin{figure}[h]
    \centering
    \includegraphics[width=\linewidth]{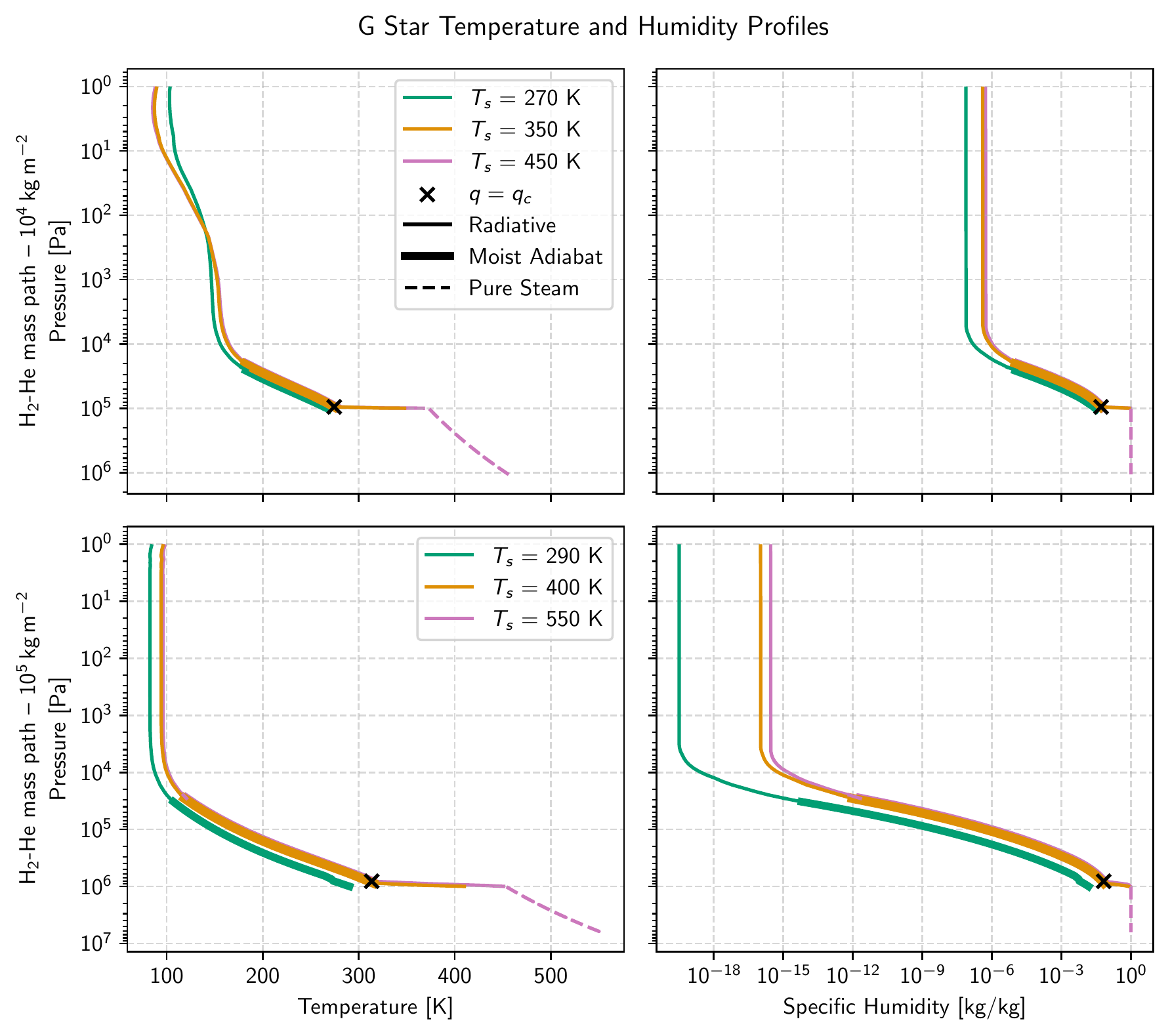}
    \caption{Same as Figure~\ref{fig:tp_mstar} but for the G star experiments.}
    \label{fig:tp_gstar}
\end{figure}

Figures~\ref{fig:tp_mstar} and~\ref{fig:tp_gstar} show sample temperature-pressure profiles and specific humidity profiles for the M star and G star experiments. At pressure levels where $q>q_c$, there is a sharp increase in lapse rate which corresponds to increased surface temperatures with respect to convecting lower atmospheres. Once the atmosphere reaches the pure steam limit in the lower atmosphere ($q=1$) the temperature profile follows a pure steam adiabat. In the M star 1 bar experiment and both G star experiments, enough SW radiation penetrates the lower atmosphere to make the radiative layers have an extremely steep lapse rate compared to both the moist adiabat and the pure steam adiabat. However, in the M star 10 bar experiment (bottom row of Figure~\ref{fig:tp_mstar}) attenuation of SW radiation results in a smoother transition between adiabatic and radiative regions.

\subsubsection{Surface Temperature vs. Instellation}
\begin{figure*}[h]
    \centering
    \includegraphics[width=\linewidth]{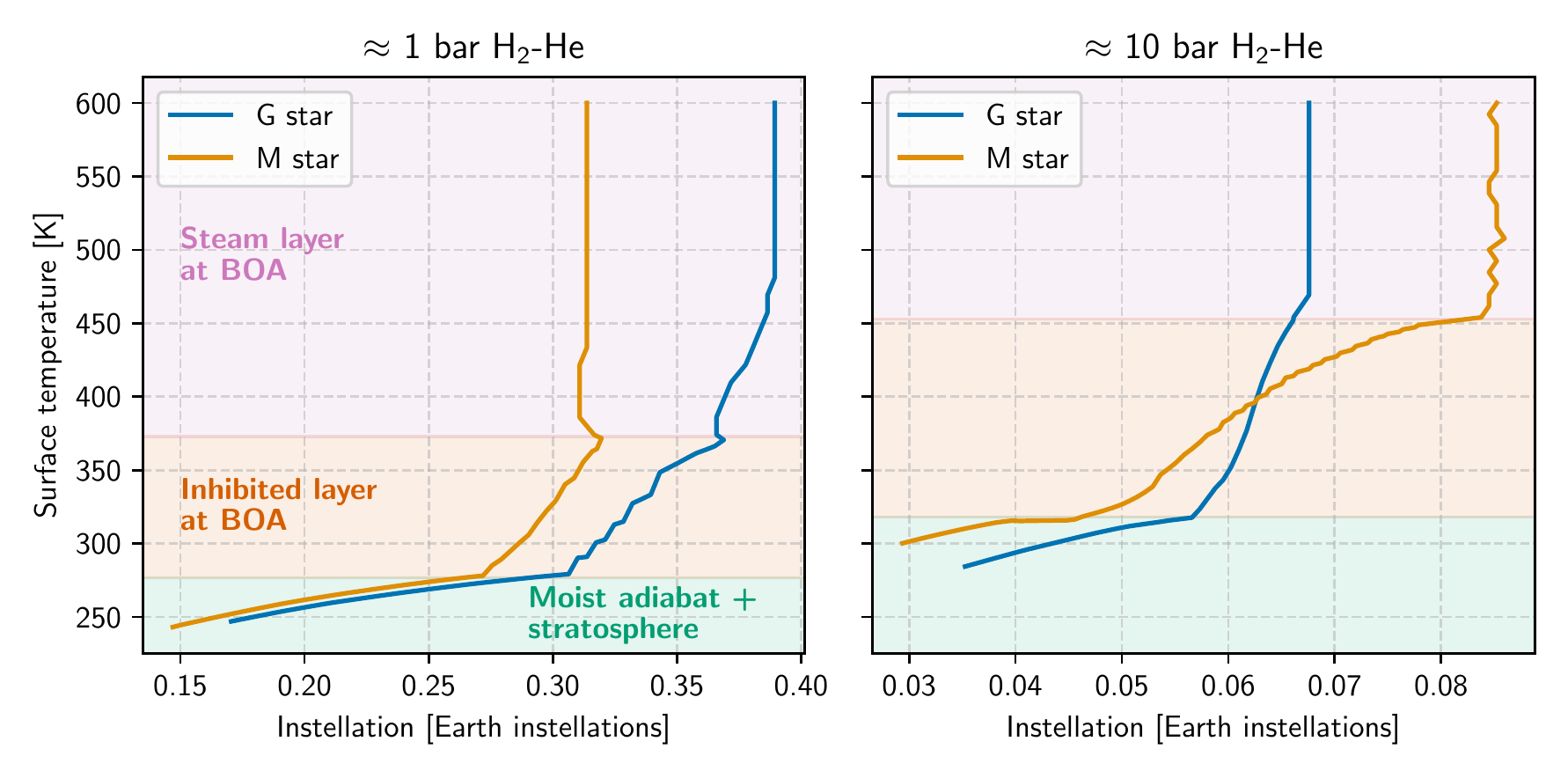}
    \caption{Surface temperature as a function of incoming instellation for different hydrogen-helium inventories. The introduction of layers with convective inhibition when the moisture content at the bottom of the atmosphere (BOA) is high enough (orange region) causes a rapid increase in surface temperature (and moisture content). Once the bottom of the atmosphere is pure steam (pink region), the atmosphere becomes optically thick and increasing surface temperature is no longer linked with an increase in cooling, causing a runaway state. \label{fig:s_vs_ts}}
\end{figure*}

Figure~\ref{fig:s_vs_ts} shows the surface temperature, $T_s$, as a function of the incoming instellation. We divide our graphs into three regions. In the region where $q(T_s, p_s) < q_c$ (green in Figure~\ref{fig:s_vs_ts}), the atmospheric $T$-$p$ structure is a moist adiabat in the lower atmopshere with a radiative stratosphere at low pressures. Once $q(T_s, p_s) > q_c$ (orange in Figure~\ref{fig:s_vs_ts}), there is a radiative layer in the lower atmosphere. The steep lapse rates in this region tend to increase the surface temperature sharply with instellation than in the lower temperature cases. Lastly, when $q(T_s,p_s) = 1$ (pink in Figure~\ref{fig:s_vs_ts}), the bottom of the atmosphere is pure steam and lies on a pure steam adiabat. At this point the atmosphere becomes optically thick at all wavelengths and the surface temperature decouples from the OLR. Further increases in the instellation cannot be compensated by additional cooling and the surface temperature increases until the ocean reservoir is depleted or the critical point of water is reached. In the latter case (applicable to sub-Neptunes with a significant water inventory), the hydrogen atmosphere is miscible with the supercritical water envelope. The possible equilibrated states of a super-runaway pure steam atmosphere are discussed in \cite{pierrehumbert2023}. The addition of hydrogen in this scenario is beyond the scope of this work.

When our initial condition is a 1 bar hydrogen-helium atmosphere, the M-star runaway limit instellation is \SI{435}{\W\per\m\squared}, and our G-star runaway limit is \SI{530}{\W\per\m\squared}. The higher runaway limit for G-stars is due to the Rayleigh scattering cross-section being larger at shorter wavelengths, which leads to an increased SW albedo for the G-star experiment where more of the instellation is at low wavelengths. 

Interesting behaviour is caused by the sudden drop in maximum possible instellation received in both cases when the atmosphere becomes pure steam (at the boundary between the pink and orange regions in Figure~\ref{fig:s_vs_ts}). For global equilibrium, we require:
\begin{equation}
\label{eq:instellation}
    S = \frac{4\text{OLR}}{1-\alpha}
\end{equation}
where $\alpha$ is the albedo. Inspecting the relevant terms, the drop in absorbed instellation at this boundary is caused by a sudden decrease in the albedo with the introduction of the steam layer. This initial drop is caused by a sudden increase in SW absorption from the sharp increase in water vapour at the bottom of the atmosphere. Less radiation reflectsfrom the surface, decreasing $\alpha$ and therefore $S$ in Equation~\ref{eq:instellation} (assuming OLR remains approximately constant). Adding more steam at the bottom of the atmosphere (increasing surface temperature) eventually causes the albedo to increase again, since the albedo of a thick pure steam layer is greater than the surface albedo, 0.12. In the G-star case, more SW radiation penetrates into the lower atmosphere than the M-star case, leading to a larger increase in albedo and hence a larger increase in instellation. 

The drop in absorbed SW radiation leads to a narrow range of instellations with multiple equilibrium surface temperatures, some of which are unstable (depending on the sign of $\dv*{S}{T_s}$, \citep{Koll2019}). In the M-star case with 1 bar \ce{H2}-\ce{He}, the drop in absorbed radiation leads to the atmosphere abruptly entering a runaway state at any instellation above 0.32 $S_0$.

With 10 bars of \ce{H2}-\ce{He} as the initial condition, we see similar overall behaviour as the 1 bar case. The boundaries between the three regimes identified above have shifted to higher surface temperatures due to the higher surface pressure. Due to the much higher CIA optical depth of the dry gas inventory compared to the 1 bar case, there is more SW absorption in these atmospheres. This mutes the varying albedo effect at the pure steam boundary described above, and leads to the surface temperature increasing monotonically with instellation towards the runaway limit. This limit is \SI{92}{\W\per\m\squared} for the G-star and \SI{116}{\W\per\m\squared} for the M-star case. We note that the M-star limit is now higher than the G-star, despite the higher albedos of the G-star irradiated atmospheres. In the M-star radiative regions at $p\approx 10$ bar, there is very little SW radiation penetrating the radiative layer. In the optically thick limit, radiative flux can be approximated as radiative diffusion \citep{Pierrehumbert2010,Heng2014}, with corresponding radiative equilibrium lapse rate:

\begin{equation}
    \label{eq:diffusion}
    \pdv{T}{p} = \frac{3}{16}\frac{\kappa}{g\sigma T^3} S_{\text{net}}
\end{equation}
where $\kappa$ is the Rosseland mean opacity and $S_{\text{net}}$ is the net SW flux penetrating the region, which is also the flux that must be carried upward through the region by radiative transfer. If $S_{\text{net}}$ is small in the radiative region (as is the case with the M star 10 bar atmospheres), then the lapse rate is also relatively small, meaning increasing $S$ has less of an effect on the surface temperature than in cases where more radiation penetrates into the deeper layers. 

The curves in Figure~\ref{fig:s_vs_ts} exhibit discrete stepping in regions where there are superadiabatic radiative layers in the atmosphere. This numerical artefact arises from the sensitivity of the surface temperature to the structure of the radiative layer. Discrete stepping is worse if the vertical resolution is low and there are few model levels in the radiative region. In this case, there is a large jump in surface temperature when a new model level becomes inhibited to convection. This behaviour motivated the use of higher vertical resolution in the lower atmosphere (discussed in Section~\ref{subsubsec:conv_adj}) which reduces the size of the jumps in $T_s$ but does not completely remove them. The structure of the stepping also changes as the instellation increment is increased or decreased, suggesting the surface temperature may be somewhat dependent on the initial state of the model (which is initialised from the final temperature profile of the previous run). We do not believe the numerical artefacts affect the conclusions of our work.

\section{Discussion}\label{sec:discussion}
\subsection{Comparison of Runaway Limits to Canonical Values}

Table~\ref{tab:limits} summarizes the runaway greenhouse instellations found in the previous section. We can compare these to the classical runaway greenhouse limit calculated for hydrogen atmospheres assumed to be on a simple moist adiabat \citep[e.g.][]{Koll2019}. Since at high $T_s$ the moist adiabat approaches the pure steam limit smoothly \citep{Koll2019}, this is the same as asking the maximum instellation a pure steam atmosphere can receive and remain in global radiative equilibrium. The maximal OLR (sometimes  called the Simpson-Nakajima limit), is approximately \SI{280}{\W\per\m\squared} for Earth's surface gravity. In global equilibrium, this must be equal to $(1-\alpha) S/4$ by Equation~\ref{eq:instellation}. Our estimation of the maximum instellation therefore depends on the calculated albedo, $\alpha$, which varies by spectral type. We calculate the runaway limit for a pure steam atmosphere to be \SI{1410}{\W\per\m\squared} and \SI{1150}{\W\per\m\squared} for our G-star and M-star cases respectively. Comparing these numbers to our model results, for 1 bar of solar \ce{H2}-\ce{He} mixture, the maximum instellation is less than half of the Simpson-Nakajima limit and is less than 10\% of the classical limit with 10 bars of \ce{H2}-\ce{He} mixture. This affects the placement of the inner-edge of the habitable zone, which is given by:
\begin{deluxetable*}{ccc}
\tablecaption{Summary of runaway instellations\label{tab:limits}}
\tablehead{\colhead{Expermiment} & \colhead{Runaway Limit[\si{\W\per\m\squared}]} ([$S_0$])&\colhead{HZ inner edge [AU]}}
\startdata
G-star, 1 bar \ce{H2}-\ce{He} & 530 (0.389)& 1.60\\
M-star, 1 bar \ce{H2}-\ce{He} & 435 (0.320)& 0.280\\
G-star, 10 bar \ce{H2}-\ce{He} & 92 (0.0676)& 3.85\\
M-star, 10 bar \ce{H2}-\ce{He}& 116 (0.0852)&0.543 \\
\hline
Simpson-Nakajima, G-star & 1410 (1.04)& 0.982\\
Simpson-Nakajima, M-star & 1150 (0.847)& 0.172\\
\enddata
\end{deluxetable*}
\begin{equation}
    d = \qty(\frac{L/L_\odot}{S/S_0})^{1/2} \text{AU}
\end{equation}
where $L/L_\odot$ is the luminosity of the star normalized by the solar value. We take $L/L_\odot = 1$ for the G-star and $\log_{10} (L/L_\odot) = -1.6$ for the M-star, the same as K2-18's luminosity \citep{Benneke2019c}. 

Our new inner-edge estimates are presented in Table~\ref{tab:limits} and are further from the host star than previous calculations. For a 10 bar dry mass inventory, our values lie outside the traditional outer edge of the habitable zone (instellations between 0.2 and 0.4 $S_0$ \citep{Kopparapu2013a}).  As a reality check, we note that the low runaway thresholds are compatible with the observed \ce{H2}-dominated outer atmosphere of Uranus, even if the interior is primarily composed of water, since the instellation of Uranus is only 0.0027 that of Earth -- more than an order of magnitude lower than the runaway threshold. Neptune is even further below the threshold. 

\subsection{Why Convective Inhibition Lowers the Runaway Limit}

\begin{figure*}[h]
    \centering
    \includegraphics[width=\linewidth]{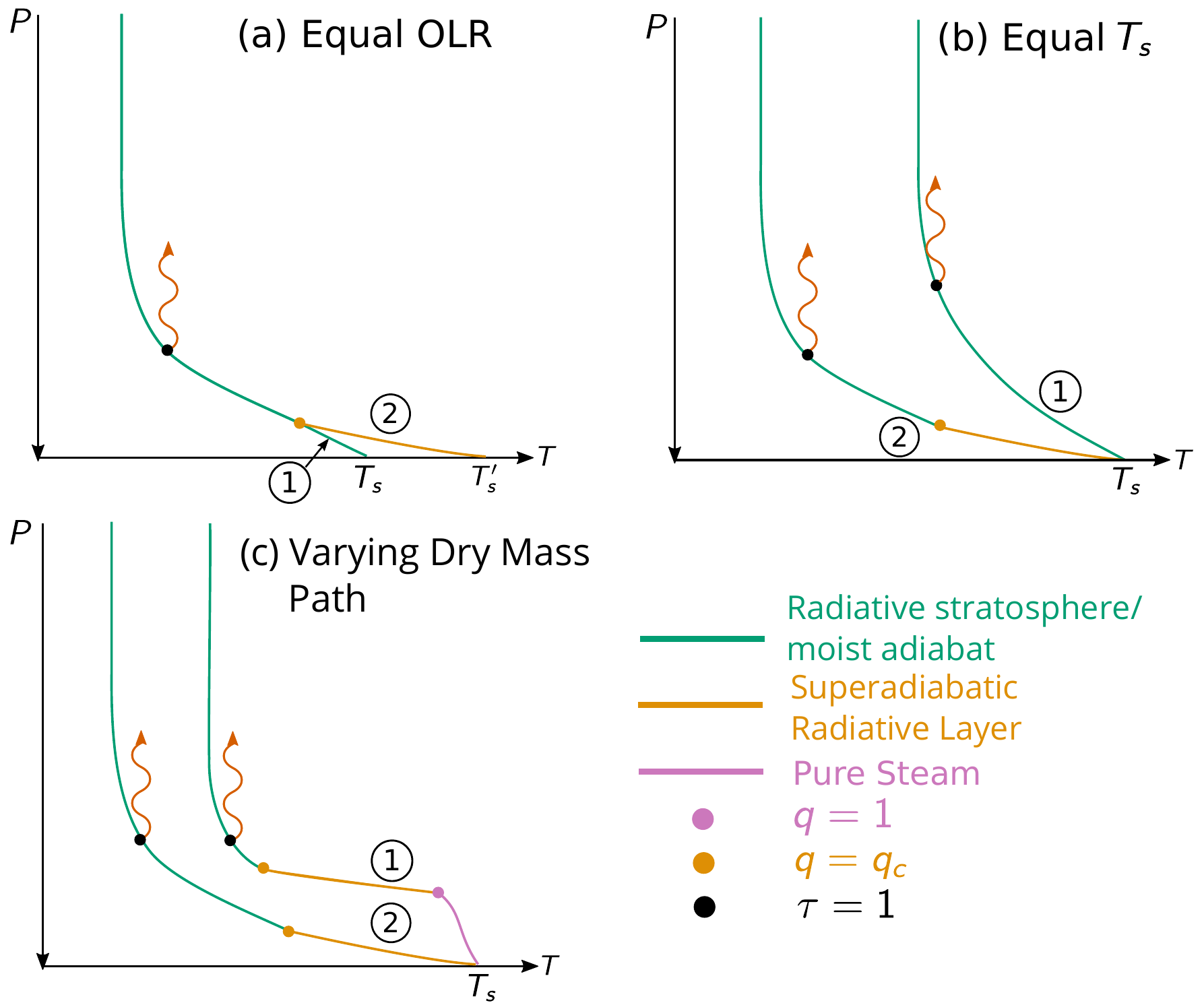}
    \caption{Three $T$-$p$ profiles demonstrating the effect of superadiabatic layers on the radiative balance of the atmosphere. (a) We compare an atmosphere with convective inhibition (2) to one without (1), assuming an identical instellation. Although the radiating temperature (and OLR) is the same between the two cases, the atmosphere with the superadiabatic layer has a higher surface temperature. (b) We again consider an atmosphere with (2) and without (1) a superadiabatic layer, starting from a fixed surface temperature. In this case, the superadiabatic layer causes the radiation temperature (and therefore OLR) of the atmosphere to decrease, which in turn reduces the maximum instellation it can receive. (c) We consider two atmospheres with inhibited layers with different dry mass paths. The pressure at $\tau=1$ is approximately constant between the two cases. The atmosphere with the greater dry mass path, (2), has a greater average lapse rate between the surface and the radiating level than (1), where the lower atmosphere is pure steam. The radiating level is therefore colder, reducing the maximum OLR of the atmosphere.}
    \label{fig:cartoon}
\end{figure*}

In this section we explore why, for a given instellation, the surface temperature is much hotter in our experiments than in traditional calculations of the inner edge of the habitable zone. Consider a planet with a given instellation, $S$. The instellation roughly sets the stratospheric temperature as $T_{\text{strat}}\sim (S(1-\alpha)/4\sigma)^{1/4}$ and the temperature of the radiating layer where the characteristic LW optical depth, $\tau$, is unity. With increasing pressure, the $T$-$p$ profile will follow a radiative layer, followed by a moist adiabatic layer until it reaches the level where $q = q_c$. In our simulations, the atmosphere then follows a radiative lapse rate, which in general is much steeper than the equivalent moist adiabatic lapse rate so long as the atmosphere is opaque enough and has enough SW radiation penetrating to that level. This increased lapse rate leads to much higher surface temperatures for equivalent levels of instellation. This effect is illustrated in Figure~\ref{fig:cartoon}(a). Once the bottom of the atmosphere becomes pure steam, the surface temperature increases steeply with small increases in instellation, since the bottom of the atmosphere becomes optically thick and decoupled from the OLR. 

Equivalently, we can imagine the temperature-pressure profile in both the classical and inhibited scenarios starting from the same surface temperature, $T_s$. If $q(T_s, p_s)> q_c$ then our modelled atmospheres will follow a steep radiative lapse rate, compared to the shallower moist adiabat. Once $q<q_c$, the inhibited atmosphere will again follow a moist adiabat (albeit one with a much steeper lapse rate than the one departing from $(T_s,p_s)$ owing to $q$ now being more dilute). The resulting upper atmosphere temperature of the inhibited atmosphere will be much lower, leading to a much lower OLR. In global equilibrium, $\text{OLR} = S(1-\alpha)/4$, and so the maximum allowed instellation for a given surface temperature will be much lower. This is illustrated in Figure~\ref{fig:cartoon}(b). 

\subsection{Why Increased Dry Mass Lowers the Runaway Limit}
\label{subsec:drymass}
It is clear from Figure~\ref{fig:s_vs_ts} that the increased dry mass path in the 10 bar \ce{H2}-\ce{He} experiments lowers the maximum instellation limit. Figure~\ref{fig:cartoon}(c) shows two different dry mass path atmospheres with the same surface temperature. The pressure level at which $\tau = 1$ is relatively constant between the two cases, because in the upper atmosphere $q<< 1$ so $\tau = 1$ when $ p(\tau=1) \approx g/\kappa_d$ where $\kappa_d$ is some characteristic grey opacity of the dry gas inventory. The temperature at which $q=q_c$ is approximately constant with temperature (see Section~\ref{subsubsec:analytic_olr}). Increasing the dry mass path of the atmosphere also increases the pressure level at which $q$ becomes unity and the atmosphere transitions to a steam layer. As seen in Figure~\ref{fig:cartoon}(c), since the pure steam layer (pink) has a much lower lapse rate than the radiative layer (orange), the average lapse rate between the surface and the $\tau=1$ level increases with increasing mass path. A higher average lapse rate decreases the radiating temperature with increasing dry mass path, leading to a drop in the OLR (and hence maximum instellation) in the runaway limit.


\subsection{Discussion of Analytic OLR}\label{subsubsec:analytic_olr}
In Appendix~\ref{sec:appendix}, we calculate the OLR from an inhibited atmosphere to be:
\begin{equation}
\label{eq:olr_anly}
   \text{OLR} = \Gamma(1 + 4\alpha) \qty(\frac{\kappa_d m_d}{\bar{\theta}})^{-4\alpha}  \qty(q_0\frac{ \kappa_v m_d }{\varepsilon \bar{\theta}})^{4/(\beta_0 - 1)}\sigma T_0^4
\end{equation}
where $T_0\approx 273$ K, $\beta_0\equiv L/(R_vT_0)$, $q_0\equiv 1/(\beta_0\varpi)$, $\alpha \equiv R_d/c_p$, $\bar{\theta} = 3/5$, $\varepsilon\equiv\mu_v/\mu_d$, $\Gamma$ is the standard gamma function and $\kappa_v$ and $\kappa_d$ are the characteristic grey opacities of the moist and dry components of the atmosphere respectively. Choosing $\kappa_v = 0.01$ \si{\m\squared\per\kg} to match the Simpson-Nakajima limit for a pure steam atmosphere, this leaves the OLR as a function of the dry opacity, $\kappa_d$ and the dry mass path, $m_d$. Table~\ref{tab:analytic} shows estimates of the analytic OLR when $\kappa_d = $ \SI{1.6e-4}{\m\squared\per\kg}, a sensible value for the dry opacity which corresponds to an \ce{H2}-\ce{He} atmosphere that becomes optically thick at approximately 0.6 bars

The term in $m_d^{-4\alpha}$ represents the total optical depth of the dry component of the atmopshere -- as this term increases, the radiating temperature drops and the OLR decreases. The second term in $m_d^{4/(\beta_0-1)}$ represents how increasing the dry mass of the atmosphere increases the temperature at which the atmosphere becomes radiative, which increases the OLR. In general, $\alpha \gg (\beta_0 - 1)^{-1}$, and we estimate $4(\beta_0-1)^{-1}-4\alpha\approx -8/7$, so the OLR decreases with dry mass, explaining the trend in Table~\ref{tab:limits}. Moreover, Table~\ref{tab:analytic} shows that the 10 bar M-star case is has a reduced dry mass compared to the equivalent G-star case, explaining its higher instellation limit since from Equation~\ref{eq:olr_anly} this atmosphere will cool more efficiently. More care could be taken to ensure that the dry mass path of the atmosphere is conserved across our different simulations. With the current model setup, this would involve iterating the surface pressure so that the integral:
\begin{equation}
    \int_0^{p_s}(1-q)\frac{\dd{p}}{g}
\end{equation}
is conserved. This was deemed too computationally expensive for the current study. Alternatively, one could implement a self-consistent moisture scheme that keeps track of the mass of the vapour phase, which would naturally conserve dry mass.

Moreover, in Appendix~\ref{sec:appendix} we show that the ratio of this OLR to the Simpson-Nakajima limit can be written as:
\begin{equation}
      \frac{\text{OLR}}{\text{OLR}_{\text{SN}}} = \frac{\Gamma(1 + 4\alpha)}{\Gamma(1 + 4/\beta_0)} \qty(q_0\frac{\kappa_v m_d}{\varepsilon\bar{\theta}})^{4/(\beta_0-1)} \qty(\frac{\kappa_d m_d}{\bar{\theta}})^{-4\alpha}  
\end{equation}
Since the ratio of gamma functions is of order unity, and $4(\beta_0-1)^{-1}\ll 1$, so long as $\kappa_d m_d/\bar{\theta}>1$, i.e. the dry inventory of the atmosphere is optically thick, we should expect the OLR to be lower than the classical Simpson-Nakajima limit.

The analytic expression provides a relatively good estimate of the OLR but decreases slightly too steeply with increasing $m_d$ -- more experiments at different dry paths would need to be run to establish the limitations of the power-law formulation. 

\begin{deluxetable*}{cccc}
\tablecaption{Comparison of model OLR with analytical calculations with a dry gas opacity of \SI{1.6e-4}{\kg\per\m\squared}\label{tab:analytic}}
\tablehead{\colhead{Expermiment} & \colhead{OLR [\si{\W\per\m\squared}]} &\colhead{Dry mass [\si{\kg\per\m\squared}]} & \colhead{Analytic OLR [\si{\W\per\m\squared}]}}
\startdata
G-star, 1 bar \ce{H2}-\ce{He} & 100& \num{9.92e4} & 108\\
M-star, 1 bar \ce{H2}-\ce{He} & 102& \num{9.01e4} &109 \\
G-star, 10 bar \ce{H2}-\ce{He} & 13.8& \num{9.74e3}&  11.8\\
M-star, 10 bar \ce{H2}-\ce{He}& 26.4 & \num{5.99e4} & 19.9\\
\enddata
\end{deluxetable*}

\subsection{What Does a Super-Runaway State Look Like?}\label{subsec:super-runaway}

Having discussed how the runaway greenhouse threshold changes for inhibited atmospheres, it is natural to wonder what a super-runaway atmosphere would look like. For Earth-like planets with a finite water reservoir, eventually the water inventory will be entirely in the atmosphere, and the lower atmosphere will not be in liquid-vapour phase equilibrium, allowing it to lie on a dry adiabat and increase its OLR \citep{Boukrouche2021}. However, for a Hycean world with an almost limitless supply of water this is not possible. As discussed in \cite{pierrehumbert2023}, a super-runaway state is likely to consist of supercritical water vapour in the lower atmosphere. Since the supercritical phase is not constrained to lie on the phase equilibrium boundary between liquid and vapour (in contrast to the condensing layers above), a deep layer of the interior heats up until the supercritical water layer penetrates to high enough altitudes that radiation to space can increase beyond the runaway limit. In the pure water case discussed in \cite{pierrehumbert2023}, this generally leaves a thin condensing region near the top of the atmosphere, but in the case with a substantial \ce{H2} layer at the top, the radiating level is in the \ce{H2} layer, so the condensing layer is eliminated entirely.  The warming proceeds until the \ce{H2} layer becomes hot enough to increase the OLR, but once supercritical water is in contact with the \ce{H2}, that layer would mix into the supercritical water and largely disappear, because \ce{H2} (and presumably also \ce{He}) is completely miscible in supercritical water \citep{soubiran2015}. A full thermal evolution model would be needed to determine how long this process would take.  A curious possibility emerges because the runaway instellation threshold with an \ce{H2} layer is considerably below the pure-steam limit. If the instellation lies between the two thresholds, once the \ce{H2} layer is diluted into the supercritical water interior, a liquid ocean could form again, forcing some \ce{H2} back into the atmosphere. One possibility is that the mixing between the layers results in just enough \ce{H2} remaining in the outer layer for radiative balance to be achieved, predicting a self-regulation in the thickness of the \ce{H2} layer. 

The scenario we have modelled in this paper corresponds to a cold-start, in which the planet begins in a sub-runaway state and then undergoes a runaway as the stellar luminosity increases. This is a possible scenario for an F or G star, but an alternate scenario for the evolution is a hot-start, in which the planet begins in a super-runaway state, either because of the heat of formation of the planet or because of intense illumination in the extended pre main-sequence stage of low mass stars. In a hot-start, the initial \ce{H2}-\ce{H2O} inventory would most likely begin in a supercritical mixed state.  While the instellation remains above the runaway threshold, a significant separated \ce{H2} layer would never form (unless \ce{H2} is a major proportion of the initial composition). Once the planet cools down enough to be sub-runaway, though, a liquid ocean will form, leading an \ce{H2} layer to effervesce out of the subcritical liquid ocean.  

\subsection{Mixing of \ce{H2}-\ce{He} into the Pure Steam Layer}

The layered structure that occurs in our model at high surface temperatures, with a pure steam layer at the bottom, is a self-consistent solution of the radiative-convective equations.  It is a peculiarity of the compositional stability criterion (Equation~\ref{eq:crit}) that when $q>q_c$, $q\rightarrow 1$ appears to be a singular limit. When $q=1$ exactly, the moist stability criterion is the usual criterion that the lapse rate be steeper than the adiabat.  However, if even an infinitesimal amount of hydrogen or helium mixes into the pure steam layer, so $q = 1-\delta$, with $\delta \ll 1$, the compositional stability criterion then nominally  applies according to which  lapse rate steeper than the moist adiabat are stable. Radiative cooling would then be expected to generate a steep radiative layer within the nearly pure steam layer, no matter how small $\delta$ may be.  However, since the stabilizing compositional buoyancy becomes exceedingly week for small $\delta$, many other mixing processes could intervene, so we find the generation of a radiative layer under these circumstances to be implausible.  We cannot rule out the possibility, though; it is a matter that will need to be resolved by future resolved-convection modelling. 

\subsection{Implications for Observations}
Our main result is that the runaway limit for sub-Neptune water worlds is greatly reduced with even 1 bar of hydrogen. Due to observational biases favouring the detection of low semi-major axis planets, most of the current observational candidates for Hycean worlds have relatively high equilibrium temperatures. From Table~\ref{tab:limits}, the inner edge of the habitable zone is around 0.28 AU for our formulation. Table~1 of \cite{madhusudhan2021} lists potential Hycean world candidates, all of which lie within our inner edge estimate and therefore would only be able to sustain a liquid water ocean if hydrogen were not detected in large abundances in the atmosphere. The well-studied sub-Neptune K2-18 b, originally thought to lie right on the inner edge of the classical habitable zone, is well beyond the inner edge by this measure. 

In this case, the planets most likely to host liquid water oceans on close-in orbits are either terrestrial planets with a higher mean-molecular weight background gas, or pure ``water-worlds" with little to no \ce{H2} envelope and an atmosphere predominantly composed of steam. The latter type of planets and their evolution have been studied previously in the highly-irradiated regime \citep{mousis2020,aguichine2021}, finding that water worlds may fit the observed mass-radius distribution of small radius planets. The super-Earth sized planet Kepler-138 d \citep{piaulet2023} is a candidate volatile-rich planet. It has a 1.5 $R_{\oplus}$ radius and low density making it not dense enough to be predominantly rocky, but also too dense to sustain a significant hydrogen envelope that would not be lost through atmospheric escape. Although this particular planet is above the runaway greenhouse instellation threshold, similar cool planets may be able to host liquid water oceans at near-Earth instellations.

The ability to observationally distinguish liquid water surfaces and super-runaway mixtures of \ce{H2}-\ce{H2O} would allow us to verify some of the predictions of this paper. Mapping the transition from sub-runaway planets to super-runaway planets as a function of instellation would give us the runaway instellation limit, which could be compared to the predictions in Table~\ref{tab:limits} to provide evidence for or against robust super-adiabatic layers. The combined non-detection of ammonia and detection of methanol in a sub-Neptune atmosphere has been proposed as a method of distinguishing a shallow water surface \citep{tsai2021}. However, this method requires around 20 transits with the James Webb Space Telescope which may prove unfeasible given time allocation restrictions. Our results also predict a sharp transition between sub-runaway atmospheres where the upper atmosphere is very dry (see Figures~\ref{fig:tp_mstar} and \ref{fig:tp_gstar}) to super-runaway atmospheres that are moist due to the mixing of \ce{H2} and supercritical \ce{H2O} \citep{pierrehumbert2023}. In contrast, the moistening of the upper atmosphere in the classical runaway greenhouse limit without convective inhibition is more smooth as instellation is increased and would occur at higher instellations. Further complications arise from the presence of non-Hycean sub-Neptunes, where the presence of water in the atmosphere is not necessarily correlated with its interior structure or the presence of a surface. More work is needed to understand how we can observationally disentangle the various possible atmospheric structures of habitable zone sub-Neptunes.

\subsection{Robustness of Calculations and Caveats}
\label{subsec:robust}

\subsubsection{Day-Night Averaging}
Our model is one-dimensional, and makes the assumption that the stellar radiation is redistributed evenly over the dayside and nightside of the planet. The superadiabatic layers in our model are sustained by the need to remove the stellar flux deposited at the surface of our model. If day-night heat redistribution is not efficient, then the nightside of these planets may be able to sustain shallower lapse rates which would aid with radiative cooling. However, for thick hydrogen atmospheres on temperate sub-Neptunes, general circulation models \citep{Charnay2021,innes2022} have shown that the combination of slow rotation rate and low mean molecular weight atmospheres produces globally weak temperature gradients thanks to dynamical redistribution of heat. This suggests heat deposited near the surface would be transported horizontally to the nightside, maintaining the steep lapse rates globally. Scaling relations for shallow atmospheres also suggest low MMW atmospheres should have efficient redistribution of heat \citep{koll2022}.

\subsubsection{Assumption of Saturation}
Our model also assumes 100\% relative humidity in the column (except above the stratospheric cold trap). One major effect of 3D dynamics is to cause subsiding regions (e.g. due to the descending branch of a Hadley-like circulation, or night side subsidence on a tidally locked exoplanet). Descending dry air causes compressional heating and undersaturation, and can be responsible for regions where the OLR is locally greater the result of a globally-averaged calculation \citep{pierrehumbert1995, Leconte2013}. If an atmosphere is undersaturated to the point it lies below the critical water vapour mixing ratio $q_c$, then superadiabatic layers responsible for surface heating may not form. Moreover, on Earth moist convection is the main mechanism by which water vapour is transported vertically in the atmosphere. Within the inhibited superadiabatic layer, mixing by convection is suppressed and our assumption of 100\% relative humidity may break down aloft. However, we note that if the near-surface layers are saturated (due to being close to the ocean surface) and the layers aloft are undersaturated, this induces an even greater mean-molecular weight gradient to stabilise the atmosphere to convection than before. Moreover, since undersaturated lofted parcels would travel on the dry adiabat, which has a steeper lapse rate than the moist adiabat, this would again help stabilise the atmosphere to convection. One could argue that decreasing relative humidity with height would affect the radiative calculations. However, from Section~\ref{subsec:drymass} and Appendix~\ref{sec:appendix} we can see that the main driver of lower cooling is the radiative effect of the dry mass of the atmosphere. A decrease in relative humidity with height would likely decrease the thickness of the superadiabatic layer, in which case the reduction in OLR may not be as severe as in the fully saturated scenario.

\subsubsection{Clouds and Hazes}
Cloud and haze opacities were not included in our model. We expect their introduction to affect our results in three main ways. Firstly, the increase in LW opacity due to cloud water or hazes would exacerbate the greenhouse effect and, if taken independently from other cloud and haze radiative feedbacks, decreases the runaway instellation. Secondly, the SW scattering properties of clouds and hazes increase the effective planetary albedo. This opposes the greenhouse effect and increases the value of the runaway instellation. Thirdly, clouds and hazes may reduce the magnitude of shortwave radiation penetrating the lower atmosphere. This would reduce the radiative lapse rates in the inhibited layers by reducing $S_{\text{net}}$ in Equation~\ref{eq:diffusion}. The reduction of radiative lapse rates will decrease the surface temperature for any given instellation, raising the runaway instellation. A similar effect was modelled in \cite{piette2020a}, who demonstrated that surface water oceans were possible on K2-18 b if the haze scattering opacity was high enough. In this case, the lower atmosphere becomes isothermal, allowing for temperate oceans at high pressures. However, their model neglected the effect of convective inhibition. Moreover, as discussed in Section~\ref{subsubsec:conv_adj}, saturated subadiabatic radiative layers are unstable to convection when $q>q_c$, implying that above the critical moisture threshold, the moist adiabat is the minimum possible lapse rate. 

If the cooling effects of clouds and hazes dominate their potential warming effect, then our runaway instellations in Table~\ref{tab:limits} are likely too pessimistic and the inner edge of the habitable zone could be at lower orbital distances. The magnitude of the cloud and haze radiative effects is likely to be strongly dependent on the particles' microphysical properties and 3D spatial distribution \citep[e.g.,][]{Yang2013,turbet2021}. These effects are beyond the scope of our simplified 1D model, though we encourage future efforts to quantify the impact of clouds and hazes on our results.

\subsubsection{Other Heat Transport Mechanisms}
We also need to consider other mechanisms which may be able to transport heat through the stabilised layers. For example, how efficient is thermal conduction at transporting flux deposited in the lower atmosphere? We can compare the efficiency of thermal conduction to radiation by comparing the thermal diffusion coefficient for conduction and radiation, as in \cite{Markham2022}. For an ideal gas, the thermal diffusivity of conduction is:
\begin{equation}
    k_{\text{cond}} = \rho \lambda c_v \sqrt{\frac{2k_B T}{\pi m}} \approx \SI{1}{\W\m\per\K}
\end{equation}
where $\lambda$ is the mean free path, $m$ is the weight of an average gas molecule, $k_B$ is the Boltzmann constant and $c_v$ its specific heat capacity at constant volume. We choose characteristic values to give an upper limit on $k_{\text{cond}}$. We use $T=300$ K, $\lambda = k_B T/(\sqrt{2} \pi d^2 p)$ with $d=290$ pm \citep[the kinetic diameter of a hydrogen molecule,][]{mehio2014} and $m=2$ amu. Density is calculated using the ideal gas law and we note the final result is independent of pressure.

The radiative diffusivity of an optically thick gas is approximately
\begin{equation}
 k_{\text{rad}} = \frac{16}{3}\frac{\sigma T^3}{\kappa\rho} \approx \SI{1e4}{\W\m\per\K}
\end{equation}
where we have used $\kappa=0.01$ \si{\kg\per\m\squared} and used the same $T$ and $\rho$ in our calculation of $k_\text{cond}$. We conclude that energy transport via radiation is much more efficient than thermal conduction. Other sources of heat transport that could be considered are advective heat transports. Although the superadiabatic regions are statically stable, eddy heat transport could play a role in transporting heat vertically, especially if the vertical wind shear is high. Moreover, we have also neglected latent heat fluxes, which could play a significant role if there is significant condensation or re-evaporation of condensates around the region of interest. Within the framework of a 1D model with no dynamics, it is very difficult to get an accurate estimate of the magnitude of these fluxes. Studying this system with a cloud-resolving model \citep[e.g.][]{lefevre2021, tan2021} is key to understanding the robustness of the superadiabatic layer against other mechanisms of heat transport. These models would also indicate the radiative effect of water clouds on the system and derived runaway greenhouse limits.

\subsubsection{Possibility of Multiple Equilibria}
Lastly, apart from in the region near the runaway limit (see Section~\ref{subsec:resultsh2h2o}), our model doesn't consider the possibility of multiple equilibrium states and hysteresis. Given that the phase structure of a water-world can change drastically on either side of the runaway limit from steam above a surface ocean to a supercritical envelope \citep{pierrehumbert2023}, there is a possibility that much warmer surface temperatures could be achieved with a similar instellation if we modelled the atmosphere with a supercritical water layer mixed with hydrogen gas. Our models represent a ``cold start", i.e. warming a planet up that initially starts with a surface water ocean. However, realistically a sub-Neptune will form hot and cool down from a state where the water is supercritical \citep{Misener2022,Markham2022}. If there are multiple equilibrium solutions, it is possible that even when the instellation lies below the runaway threshold, the water will still be in a supercritical state. Moreover, in our model we have neglected the effect of internal heating from the residual heat of formation or tidal heating. Given that our model only requires $\approx \SI{1}{\W\per\m\squared}$ of solar flux penetrating the lower layers to drive steep superadiabatic lapse rates, a similar level of internal flux could equally sustain very high surface temperatures in the absence of a significant instellation. 

\section{Conclusions}

The aim of this study was to determine the sustainabilty of a liquid water ocean on a Hycean world with a significant \ce{H2}-\ce{He} inventory. Our major findings are:
\begin{enumerate}
    \item Neglecting water vapour feedbacks, 10-20 bars of solar \ce{H2}-\ce{He} mixture will drive a surface ocean supercritical when forced with solar instellation. A planet receiveing 10 times solar instellation would have to have less than 1 bar of hydrogen to sustain a liquid water ocean.
    \item Including water vapour feedbacks, the presence of superadiabatic layers where convection is inhibited in the lower atmosphere reduces the runaway greenhouse instellation limit from the Simpson-Nakajima limit significantly. For a solar \ce{H2}-\ce{He} inventory of around \SI{1e4}{\kg\per\m\squared}, the runaway greenhouse limit to an instellation of approximately \SI{530}{\W\per\m\squared} for a G-star and \SI{435}{\W\per\m\squared} for an M-star. This reduces further to around \SI{100}{\W\per\m\squared} for an \ce{H2}-\ce{He} inventory of around \SI{1e5}{\kg\per\m\squared}. 
    \item The reduced instellation limits correspond to moving the inner edge of the habitable zone to around 1.6 AU (3.85 AU) for a planet orbiting a G-star with 1 bar (10 bar) of \ce{H2}-\ce{He} and equivalently 0.280 AU (0.543 AU) for a planet orbiting an M-star (c.f. 0.982 AU and 0.172 AU for a G-star and M-star respectively from previous models).
    \item Analytical models of the OLR show the key parameter responsible for the reduction in the OLR is the total optical depth of the dry inventory, given that steep superadiabatic lapse rates in the inhibited layers dry the atmosphere aloft. A higher dry optical depth reduces the radiating temperature of the atmosphere and caps the maximum cooling from a \ce{H2}-\ce{H2O} atmosphere. If we model the atmosphere as having a constant, gray opacity for the dry gases, then the limiting OLR scales roughly as the inverse of the dry mass path.
    \item Our results suggest that most of the current Hycean world targets are within the inner limit of the habitable zone and unlikely to host liquid water oceans. The most promising targets for observing a liquid water ocean on a close-in orbit are therefore traditional terrestrial-like planets with a high-mean molecular weight background atmosphere or ``water worlds" with negligible \ce{H2}-\ce{He} envelopes.
\end{enumerate}

We conclude by encouraging the use of 3D cloud resolving models to study the robustness of the inhibited, superadiabatic radiative layers to 3D dynamics and other sources of heat flux.

\begin{acknowledgments}
This paper is supported by funding from the European Research Council (ERC) under the European Union's Horizon 2020 research and innovation programme (Grant agreement No. 740963). S.-M.T. thanks Matej Malik and Daniel Kitzmann for the discussion on the water continuum absorption. S.-M.T. acknowledges support from the University of California at Riverside and NASA Exobiology grant 80NSSC20K1437.
\end{acknowledgments}

\appendix 

\section{Calculation of radiative fluxes}
\label{sec:app_radflux}
We calculate the radiative fluxes using the SOCRATES code, based on \cite{Edwards1996}. SOCRATES uses the correlated-$k$ method to efficiently calculate two-stream fluxes in both the longwave (LW) and shortwave (SW) regions of the spectrum. Gaseous overlap is treated with the equivalent extinction method with resorting and rebinning \citep{Lacis1991}. In Section~\ref{sec:h2he}, we calculate LW fluxes in 300 bands equally spaced in wavenumber between \SI{1}{\per\cm} and \SI{5000}{\per\cm}. In the SW region, we perform two separate calculations for a G-type star (using the \cite{Lean2012} solar spectrum) and an M-type star (calculated using PHOENIX \citep{Husser2013} for a \SI{3500}{K} star with $\log(g) = 5.0$ and solar metallicities and alpha element abundances). In Section~\ref{sec:h2he} we use 300 bands equally spaced in wavenumber in each of the LW and SW regions of the spectrum. For the calculations in Section~\ref{sec:h2h2o} we use 30 bands in each of the LW and SW regions to speed the convergence of the numerical iteration. We found no significant differences in the temperature-pressure profiles in benchmark models run with 300 bands in each region compared to 30 bands, justifying this reduction in wavenumber resolution. For the M-star we use the range \SI{250}{\per\cm} to \SI{40000}{\per\cm} and for the G-star we use \SI{250}{\per\cm} to \SI{50000}{\per\cm}. We consider the collision-induced absorption (CIA) due to \ce{H2}-\ce{H2} and \ce{H2}-\ce{He} interactions as the main sources of absorption. We calculate \ce{H2}-\ce{He} opacities using the HITRAN database \citep{Karman2019} and use the HITRAN database with additional data from \cite{Borysow2002} for the calculation of \ce{H2}-\ce{H2} opacities. In the SW calculation we include the effects of Rayleigh scattering by both hydrogen and helium, calcualting the cross sections using fits of refractive indices taken from \cite{Peck1977} and \cite{CliveCuthbertson1932} respectively. A SW surface albedo of 0.12 was specified \citep{goldblatt2013} and the surface temperature was assumed to be identical to the temperature of the lowest layer of the atmosphere.

Water vapour absorption $k$-coefficients are calculated using HITRAN data \citep{Karman2019} and continuum absorption is calculated with the MT\_CKD model \citep{Mlawer2012}. The refractive index of water used to calculate the Rayleigh scattering coefficients was taken from \cite{Ciddor1996}.
\section{Analytic OLR calculations}
\label{sec:appendix}
We can understand our results better by constructing a simple grey analytic framework for calculating the maximum OLR of our model atmospheres.

In the optically thick limit ($\tau\gg 1$), the OLR of a grey atmosphere is given by:
\begin{equation}
\label{eq:OLR}
    \text{OLR} \approx \int_0^{\infty} \sigma T(\tau)^4 e^{-\tau} \dd{\tau}
\end{equation}
To calculate this integral, we require the temperature profile $T(\tau)$ and the optical depth:
\begin{equation}
    \tau(p) = \int_0^{p}\frac{\dd{p'}}{\bar{\theta}g}\qty(\kappa_d (1-q) + q\kappa_v)
\end{equation}
where $\kappa_d$ and $\kappa_v$ are characteristic grey opacities for the dry and condensible phases of the atmosphere respectively and $\bar{\theta}$ is the average zenith angle, accounting for the angular distribution of the LW radiation. To simplify the problem, we split our atmospheres into two regions separated at $q = q_c = 1/(\beta\varpi)$, since we expect the lapse rates in these two regions to be very different. We assume that the moisture at this level is dilute such that we can make the approximation:
\begin{equation}
\label{eq:qdil}
    q \approx \varepsilon \frac{p_{\text{sat}}}{p}
\end{equation}
This approximation is approximately valid if $q_c<0.1$, i.e. $T < 0.1 L\varpi/R_v  = 480$ K assuming $\beta = L/R_vT$. We can verify that this is the case in all of our simulations. We then approximate $p_{\text{sat}}(T)$ in the same way as \cite{Koll2019}, writing:
\begin{equation}
    p_{\text{sat}}(T) = p_0\qty(\frac{T}{T_0})^{\beta_0}
\end{equation}
where $\beta_0\equiv L/(R_vT_0)$ and $(p_0,T_0)$ is some reference point on the vapour-liquid phase curve of water chosen to be close to our region of interest. Like in \cite{Koll2019}, we will choose $p_0$ as the pressure where a pure steam atmosphere has unity optical thickness:
\begin{equation}
p_0 = \frac{g\bar{\theta}}{\kappa_v}    
\end{equation}
This gives $(p_0,T_0) = (588.6~\text{Pa}, 272.6~\text{K})$ for $g = 9.81$ \si{\m\per\second\squared}, $\kappa_v = 0.01$ \si{\m\squared\per\kg},  $\bar{\theta} = 3/5$ -  remarkably close to the triple point of water vapour. 
If Equation~\ref{eq:qdil} holds, then the temperature and pressure $(T_*,p_*)$ at which $q = q_c$ satisfies:
\begin{align}
    \varepsilon\frac{p_0}{p_*}\qty(\frac{T_*}{T_0})^{\beta_0} &= \frac{R_v T_0}{p}\\
    T_* &= T_0 \qty(\frac{1}{\beta_0\varpi}\frac{p_*}{\varepsilon p_0})^{1/(\beta_0-1)} \label{eq:Tstar}
\end{align}

By noting that $\beta_0\gg 1$ in our range of $T_0$ \citep{Koll2019}, we can readily verify our dilute approxmation:

\begin{equation}
    \frac{T_*}{L\varpi/R_v}\approx \frac{1}{\beta_0\varpi} = 0.06
\end{equation}
for $T_0 = 272.6$ K. We will also define $q_0 = 1/(\beta_0\varpi)$ as the moisture inhibition threshhold at $T = T_0$

We then want to relate our pressure $p_*$ to the dry mass path. We note that since $q_c$ is relatively dilute, the atmosphere will quickly dry on the moist adiabat extending upwards from this point, leaving most of the upper atmosphere dry. For a dry mass path $m_d$, we can then write $p_*\approx m_d g$. 

We assume that the temperature structure of the upper atmosphere ($p<p_*$) is approximately a dry adiabat emanating from $(p_*,T_*)$. This neglects the effect of moisture on the lapse rate, which can be significant but quickly leads to intractable solutions since lapse rate $\dv*{T}{p}$ depends on $q(p,T)$. This dry adiabat has the form:

\begin{equation}
    T = T_*\qty(\frac{p}{p_*})^{\alpha}, \quad \alpha \equiv \frac{R_d}{c_p}
\end{equation}

This allows us to write an equation for how $\tau$ varies with $\ln T$:

\begin{align}
    \dv{\tau}{\ln T} &= \qty(\dv{\ln T}{\ln p})^{-1}p\dv{\tau}{p}\\
                     &= \alpha^{-1}\qty[ \frac{\kappa_d p}{\bar{\theta}g} + (\kappa_v - \kappa_d)\frac{\varepsilon p_{\text{sat}}}{\bar{\theta}g}]\\
                     & = \alpha^{-1}\qty[ \frac{\kappa_d p_*}{\bar{\theta}g} \qty(\frac{T}{T_*})^{1/\alpha} + (\kappa_v-\kappa_d)\frac{\varepsilon p_{0}}{\bar{\theta}g} \qty(\frac{T}{T_0})^{\beta_0}]
\end{align}
Integrating this relation and letting $\tau = 0$ when $(T,p) = (0,0)$ yields:
\begin{equation}
\label{eq:tau}
    \tau = \alpha^{-1} \qty[\frac{\alpha \kappa_d p}{\bar{\theta}g} + (\kappa_v - \kappa_d) \frac{p}{\beta_0 \bar{\theta}g} q]
\end{equation}
To calculate the OLR, ideally we would invert Equation~\ref{eq:tau} to find $T(\tau)$ in Equation~\ref{eq:OLR}. However, due to the mixed powers of $T$ in Equation~\ref{eq:tau} (one term in $T^{1/\alpha}$ and one in $T^{\beta_0}$), this cannot be done analytically. To proceed, we compare the magnitude of the two terms and argue that the second term can be neglected so long as:
\begin{equation}
\label{eq:qlimit}
    q \ll \frac{\beta_0\alpha}{\kappa_v/\kappa_d - 1}
\end{equation}
Let us take characteristic values of $\beta_0$ at $T_0 = 273$ K, $\alpha \approx 2/7$, and $\kappa_v = 0.01$ \si{\m\squared\per\kg} and $\kappa_d =$ \SI{1.6e-4}{\m\squared\per\kg} (this value of $\kappa_d$ yields good agreement between our final analytical OLR and simulations and is consistent with \ce{H2} becoming optically thick in the infra-red between 0.1 and 1 bar). In this case, Equation~\ref{eq:qlimit} gives:

\begin{equation}
    q \ll 0.1
\end{equation}
We assert this to be true around the $\tau \sim 1$ region of the atmosphere, even if it doesn't strictly hold at $p = p_*$.

We then have:

\begin{equation}
\label{eq:kappastar}
T = T_* \qty(\frac{\tau}{\tau_*})^{\alpha}, \quad \tau_*\equiv \frac{\kappa_d p_*}{\bar{\theta}g} = \frac{\kappa_d m_d}{\bar{\theta}}
\end{equation}
which when inserted into~\ref{eq:OLR} yields
\begin{equation}
\text{OLR}  =  \Gamma(1 + 4\alpha) \tau_*^{-4\alpha}\sigma T_*^4
\end{equation}
where $\Gamma$ is the standard gamma function. Expanding $T_*$ from Equation~\ref{eq:Tstar} and $\tau_*$ from Equation~\ref{eq:kappastar} yields:

\begin{equation}
\label{eq:OLRestim}
    \text{OLR} = \Gamma(1 + 4\alpha) \qty(\frac{\kappa_d m_d}{\bar{\theta}})^{-4\alpha}  \qty(q_0\frac{ \kappa_v m_d }{\varepsilon \bar{\theta}})^{4/(\beta_0 - 1)}\sigma T_0^4
\end{equation}
We immediately see that this OLR limit does not depend on the surface temperature, which is characteristic of a runaway greenhouse atmosphere. The first bracket corresponding to $\tau_*$ represents the effect of increasing the dry opacity of the atmosphere. This shifts the radiating temperature up the adiabat with exponent $\alpha$, and therefore reduces the OLR of the atmosphere. The second bracket traces back to Equation~\ref{eq:Tstar} and represents how increasing the dry mass of the atmosphere increases the temperature at which the atmosphere becomes inhibited, $T_*$. Increasing the base temperature of the adiabat is associated with an increase in the OLR, albeit with a weak dependence of $m_d^{4/(\beta_0-1)}$. We can see that the OLR depends on the dry mass path $m_d$ with an exponent of:
\begin{equation}
    4(\beta_0 - 1)^{-1} - 4\alpha \approx - 4 \alpha
\end{equation}
since $\beta_0\gg 1$. For a diatomic ideal gas $-4\alpha \approx -8/7$, so we would expect the OLR to drop of sharply with dry mass path. 

We can compare Equation~\ref{eq:OLRestim} with Equation~24 in \citep{Koll2019}, for a pure steam atmosphere, which was an estimate of the classical Simpson-Nakajima limit:

\begin{equation}
    \text{OLR}_{\text{SN}} =  \Gamma(1 + 4/\beta_0) \qty(\frac{\kappa_v p_0}{\bar{\theta}g})^{-4/\beta_0}\sigma T_0^4 = \Gamma(1 + 4/\beta_0) \sigma T_0^4
\end{equation}
Taking the ratio of this equation and Equation~\ref{eq:OLRestim} gives:
\begin{equation}
    \frac{\text{OLR}}{\text{OLR}_{\text{SN}}} = \frac{\Gamma(1 + 4\alpha)}{\Gamma(1 + 4/\beta_0)} \qty(q_0\frac{\kappa_v m_d}{\varepsilon\bar{\theta}})^{4/(\beta_0-1)} \qty(\frac{\kappa_d m_d}{\bar{\theta}})^{-4\alpha}
\end{equation}

In general, since $4/(\beta_0 - 1)\ll 1$ and the gamma function ratio is of order unity, so long as $\kappa_d m_d/\bar{\theta} >1 $ (i.e. the dry mass of the atmosphere is optically thick), the OLR limit will be lower than the classical Simpson-Nakajima limit.

Lastly, if we note that if one were to take the opposite limit of Equation~\ref{eq:qlimit} (i.e. the moist optical depth at $\tau\sim 1$ dominates the OLR, we would find:
\begin{equation}
    \text{OLR} = \Gamma(1 + 4/\beta_0) \qty(\frac{c_p}{LT_0})^{-4/\beta_0}\sigma T_0^4  
\end{equation}
which is identical to the ``dilute limit" found in \cite{Koll2019}. This limit does not depend on the dry mass path and is only moderately lower than the Simpson-Nakajima limit. Since our results vary greatly with dry mass path and are much lower than this limit, this should reassure us that Equation~\ref{eq:qlimit} is a good assumption.
\bibliography{references}
\end{document}